# Evolution from Topological Dirac Metal to Flat-band-Induced Antiferromagnet in Layered K$_x$Ni$_4$S$_2$ (0⩽x⩽1)

Hengdi Zhao,[1,†] Xiuquan Zhou,[1,2,†] Hyowon Park,[1,3] Tianqi Deng,[4,5] Brandon Wilfong,[6,7] Alann P. Au, II,[8] Samuel E. Pate,[1,9] Craig M. Brown,[10] Hui Wu,[10] Tushar Bhowmick,[11] Tessa McNamee,[11] Ravhi Kumar,[3] Yu-Sheng Chen,[12] Zhi-Li Xiao,[1,9] Russell Hemley,[13] Weizhao Cai,[11] Shanti Deemyad,[11] Duck-Young Chung,[1] Stephan Rosenkranz,[1] and Mercouri G. Kanatzidis*[1,8,14]

[1]Materials Science Division, Argonne National Laboratory, Lemont, IL 60439, USA
[2]Department of Chemistry, Georgetown University, Washington D.C. 20057, USA
[3]Department of Physics, University of Illinois at Chicago, Chicago, Illinois 60607, USA
[4]State Key Laboratory of Silicon and Advanced Semiconductor Materials, School of Materials Science and Engineering, Zhejiang University, Hangzhou, China.
[5]Institute of Advanced Semiconductors and Zhejiang Provincial Key Laboratory of Power Semiconductor Materials and Devices, ZJU-Hangzhou Global Scientific and Technological Innovation Center, Zhejiang University, Hangzhou, China.
[6]Department of Chemistry, Johns Hopkins University, 3400 N. Charles Street, Baltimore, MD 21218, United States of America
[7]Institute for Quantum Matter, William H. Miller III Department of Physics and Astronomy, Johns Hopkins University, 3400 N. Charles Street, Baltimore, MD 21218, United States of America
[8]Department of Chemistry, Northwestern University, Evanston, Illinois 60208, USA
[9]Department of Physics, Northern Illinois University, DeKalb, Illinois 60115, USA
[10]NIST Center for Neutron Research, National Institute of Standards and Technology, Gaithersburg, MD 20899
[11]Department of Physics and Astronomy, University of Utah, Salt Lake City, Utah 84112, United States
[12]NSF's ChemMatCARS, the University of Chicago, Lemont, Illinois 60439, USA
[13]Departments of Physics, Chemistry, and Earth and Environmental Sciences, University of Illinois at Chicago, Chicago, Illinois 60607, USA
[14]Lead contact

*corresponding author. m-kanatzidis@northwestern.edu

†These authors contributed equally to this work.

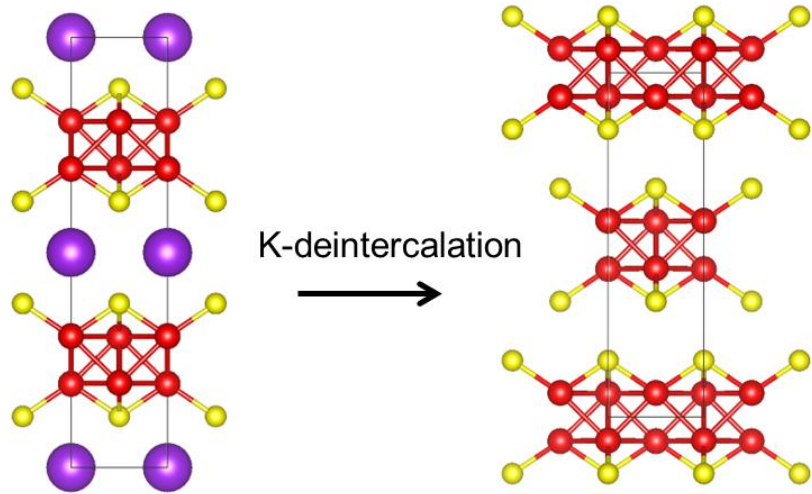

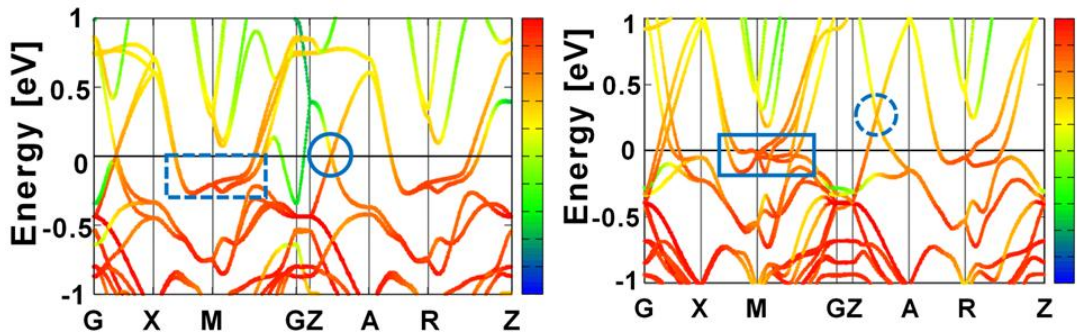

Dirac cones with massless fermions and flat bands featuring massive electrons coincide in $K_xNi_4S_2$ ($0 \leqslant x \leqslant 1$) and can be accessed selectively by continuous tuning of the Fermi level via topochemical potassium deintercalation. The ability to traverse between a topological Dirac metal and a flat-band-induced antiferromagnet within a single crystalline system, without relying on artificial lattice engineering or exotic structures like Kagome or honeycomb lattices, represents an experimentally validated platform for exploring emergent correlated phenomena.

**Progress and Potential Statement**

Dirac materials and flat band systems, each possessing distinct electronic structures, have captivated a wide range of scientific communities for their potential to host diverse emerging phenomena. In particular, a tunable ground state featuring a Fermi surface dominated by massive fermions from the flat band and massless fermions from the Dirac cone offers an ideal platform to study the interplay between these emerging phenomena. Despite great interest in such systems, materials with coexisting Dirac cones and flat bands are rare, relying on artificial lattice engineering such as twisted bilayer graphene, or exotic structures like Kagome or honeycomb lattices. In addition, the lack of an effective method for tuning the Fermi level poses another challenge. Here, we report a layered quantum material, $K_xNi_4S_2$ ($0 \leqslant x \leqslant 1$), that simultaneously hosts both flat bands and Dirac cones at distinct energies without involving the typical Kagome or honeycomb lattice. Our molecular orbital bonding analysis suggests that the Ni-Ni bonding exclusively hosted by $K_xNi_4S_2$ plays a vital role in the formation of Dirac cones. Notably, the K-content can be controlled through the K-deintercalation process, enabling the long-sought effective method of wide-range tuning of the Fermi level. With first principles calculation and experimental confirmation, we demonstrate the versatile ground state that can be fine-tuned through the K-deintercalation process, from a non-magnetic topological Dirac metal ($KNi_4S_2$, x = 1), to a flat-band induced antiferromagnet ($Ni_2S$, x = 0). The $K_xNi_4S_2$ ($0 \leqslant x \leqslant 1$) system offers an experimentally validated, versatile platform for exploring emerging phenomena from massless Dirac fermions, flat-band heavy electrons, and the interplay between them. This ex-situ topochemical K-deintercalation study also establishes a highly tunable ground state, demonstrating a viable pathway for in-situ control of quantum materials that can switch between Dirac-cone- and flat-band-dominated states via electrochemical intercalation and deintercalation.

**Highlights**
- Coexistence of flat bands and Dirac cones without Kagome/honeycomb lattices
- Continuously tunable Fermi surface through topochemical K-deintercalation
- Switchable ground state between Dirac-cone- and flat-band-dominated regimes
- Establishment of a new material design paradigm for correlated topological systems

**Summary**

Condensed matter systems with coexisting Dirac cones and flat bands, and a switchable control between them within a single system, are desirable but remarkably uncommon. Here we report a layered quantum material system, $K_xNi_4S_2$ ($0 \leq x \leq 1$), that simultaneously hosts both characteristics without involving typical Kagome/honeycomb lattices. Enabled by a topochemical K-deintercalation process, the Fermi surface can be fine-tuned continuously over a wide range of energies. Consequently, a non-magnetic Dirac-metal state with a topological nontrivial $Z_2$ index of 1;(000), supported by first-principles calculations and high mobility up to 1471 $cm^2V^{-1}s^{-1}$, is observed on the K-rich x = 1 side, whereas a flat-band induced antiferromagnetic state with $T_N$ up to 10.1 K emerges as K-content approaches 0. The $K_xNi_4S_2$ system offers a versatile platform for exploring emerging phenomena and underscores a viable pathway for in-situ control of quantum materials dominated by Dirac cones, flat bands, and their interplay.

**Keywords:** Dirac metal, non-trivial band topology, flat-band induced magnetism, topochemical deintercalation, non-Fermi liquids, subsulfide, layered materials, strongly correlated electrons

**Introduction**

Flat bands give rise to massive electrons with a large density of states (DOS), fostering a range of strongly correlated phenomena, including superconductivity,[1-4] magnetism,[5-10] heavy fermion behavior,[11-14] charge density waves,[15] fractional quantum anomalous Hall effect,[16] exceptionally large Seebeck coefficients,[17-19] etc. In contrast, Dirac cones feature linear band dispersions, leading to high carrier mobilities and fractional quantum Hall effects[20,21] due to the vanishing effective mass and contributing minimal DOS to the Fermi surface when the Dirac cone is right at the Fermi level. Dirac cones are typically observed in graphene[22-24] and analogous materials featuring the honeycomb lattice, such as silicene,[25] germanene,[26] and stanene,[27] and topological materials.[28-32]

Materials that exhibit a trivial progression of ground states through systematic doping, where each state is only marginally different from the next, are relatively common. In contrast, materials that can be tuned to access two or more fundamentally distinct electronic ground states, such as switching between a flat-band system with massive carriers and one hosting massless Dirac fermions are rare and scientifically far more valuable. These systems, where external perturbations can induce qualitative, not merely incremental, transformations in the Fermi surface, present exceptional opportunities for discovering and controlling emergent quantum phenomena. Such coexisting and switchable states have been predicted for cold atoms in optical lattices[33,34] and experimentally realized for polaritons in a photonic lattice in the semiconductor microcavity.[35] However, the experimental realization in a condensed matter system is rare, chiefly because of limited understanding of how to design a system hosting both characteristics simultaneously, and subsequently how to switch the ground state between them by manipulating the Fermi level. The artificial structure of twisted bilayer graphene offers an experimental demonstration in the condensed matter system, showing the Fermi surface dominated by flat bands near the magic-angle twisting angle, leading to multiple superconducting domes separated by Mott insulating states,[1-4] and by Dirac cones at different twisting angles.[36] As for bulk material systems, the kagome architecture naturally hosts both flat bands and Dirac cones, as depicted in the nearest-neighbor tight-binding model.[37,38] Despite the rich phenomenology predicted in the kagome materials, the experimental realization of highly tunable electronic states between Dirac cones and flat bands in bulk kagome materials remains a challenge.[39,40] Consequently, most experimental and theoretical efforts focus on only one of such characteristics, for example, flat-band induced correlated phenomena such as magnetism,[41-44] non-trivial band topology from Dirac cones,[42,45] or Van Hove singularity (VHS) driven electronic instabilities,[46-51] depending on the specific electronic band fillings. Recently, a tunable Fermi level was demonstrated between $Ni_3In$ and $Ni_3Sn$. However, the replacement of In with Sn shifts the Fermi level further away from all the interesting characteristics.[52] Nevertheless, the ability to finely tune the electronic structure from a Dirac-cone-dominated state to one governed by flat bands within a single system, to our knowledge, has yet to be demonstrated in bulk quantum materials.

Here, we show that the system $K_xNi_4S_2$ ($0 \leq x \leq 1$)[53] simultaneously hosts flat bands and Dirac cones at distinct energies without involving the kagome or honeycomb lattices. Our molecular orbital (MO) bonding analysis suggests that the Ni-Ni bonding exclusively hosted by $K_xNi_4S_2$ plays a vital role in the formation of Dirac cones. This finding aligns with a symmetry-based analysis demonstrating that the square lattice involving direct Ni-Ni bonding with dominating $d_{x^2-y^2}$ orbitals has the potential to host Dirac cones.[54] On the other hand, the flat band and its associated magnetism originate from the dominant Ni-Ni-$d_{x^2-y^2}$ orbital bonding[53] and Ni-S bonding[55] close to the Fermi level.

Moreover, the potassium content can be controlled through a topotactic K-deintercalation process, providing an effective method to adjust the Fermi level from the higher lying Dirac state to the lower lying flat band state. Both first-principles calculations and experimental evidence validate that the Fermi level and the resultant electronic states of this material can be fine-tuned from the Dirac-cone-dominating state ($KNi_4S_2$, x = 1) to the flat-band-dominating state ($K_0Ni_4S_2$, x = 0). In addition, a non-Fermi liquid behavior with a universal linear-temperature-dependent electrical resistivity is observed across all K-levels, suggesting the strange metal behavior. Such non-Fermi liquid behavior is often associated with emerging collective excitations near a quantum critical point, as seen in high $T_c$-cuprates,[56] iron-based superconductors,[57] and heavy-fermion metals.[12,14,58] Similarly, such linear or sublinear transports are also widely observed in Kagome metal systems,[59] which can be described by a semiclassical two-pocket model highlighting the interplay between fast electrons from Dirac cones and weak dispersive electrons from flat bands or Van Hove singularities.[60] Therefore, the presence of strange metal behavior indicates that strong correlations between the Dirac cones and flat bands are present across the whole series of $K_xNi_4S_2$ ($0 \leq x \leq 1$), whose ground state alternates between a non-magnetic Dirac metal on the x = 1 side and flat-bands induced antiferromagnetic metal on the x = 0 side through the K-deintercalation process.

**Origin of Coexisting Dirac Cones and Flat Bands**

KNi$_4$S$_2$ (x = 1) crystallizes in the CeRe$_4$Si$_2$-type[61] with space group *Cmmm* (**Table.S1-S2**), featuring two Ni sheets sandwiched between two S sheets (**Fig.1a**), and unconventionally low formal oxidation states for nickel of Ni$^{0.75+}$. The potassium atoms in KNi$_4$S$_2$ can be deintercalated topotactically in various degrees until the formation of the completely K-free Ni$_2$S (x = 0)[53] with a formal oxidation state of Ni$^{1+}$. Ni$_2$S crystallizes in the space group *Cmcm*, exhibiting the same structural features of KNi$_4$S$_2$ but having van der Waals gaps between the layers (**Fig.1b**). There are extensive Ni-Ni bondings in the structure because of the low valent electron-rich character on these atoms. Consequently, certain Ni atoms can be bonded to nine other Ni atoms to form a Ni$_9$ cluster. (**Fig.1c**)

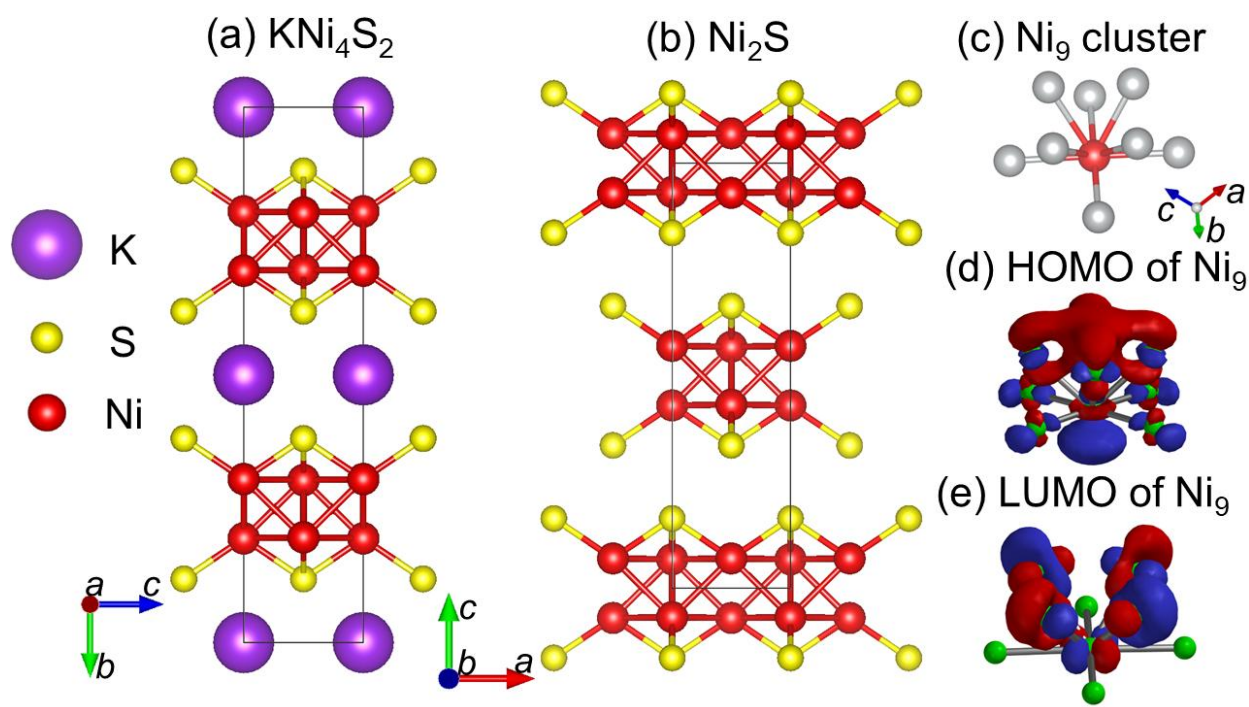

**Figure 1. Crystal structures and origin of Dirac cones.** (**a**) KNi$_4$S$_2$ (x = 1, space group *Cmmm*) and (**b**) Ni$_2$S (x = 0, space group *Cmcm*). The atoms of K, Ni, and S are depicted in purple, red, and yellow colors, respectively. (**c**) A Ni$_9$ cluster with red for the center Ni atom and grey for the surrounding Ni atoms, highlighting the Ni-Ni bonding in K$_x$Ni$_4$S$_2$ and the molecular orbitals of its (**d**) highest occupied molecular orbital (HOMO) and (**e**) lowest occupied molecular orbital (LUMO). Note that the red and blue colors denote + and − signs of MO, respectively.

To gain deeper insights into such Ni-Ni bonding and its implications, we performed density functional theory (DFT) calculations on an isolated Ni$_9$ cluster (**Fig.1c**). Its highest occupied molecular orbital (HOMO) (**Fig.1d**) shows strong Ni-$d_{z^2}$ character at the center and $d_{x^2-y^2}$ for the surrounding Ni. Conversely, the lowest unoccupied molecular orbital (LUMO) (**Fig.1e**) reveals the opposite trend, showing strong Ni-$d_{x^2-y^2}$ character at the center and $d_{z^2}$ for the surrounding Ni. The protruding $d_{z^2}$ orbitals perpendicular to the square net resemble the $p_z$ orbitals in graphene, which is essential for the creation of Dirac cones in the honeycomb lattice. A two-band symmetry analysis for the square lattice also suggested the possible formation of Dirac cones using a $s/d_{x^2-y^2}$ model,[54] where the $s$ band can be replaced by $d_{z^2}$ for the same $a_{1g}$ symmetry. This bonding analysis suggests pivotal roles played by the unique Ni-Ni interactions in the formation of Dirac cones hosted by the K$_x$Ni$_4$S$_2$ system.

**Tuning the Fermi Level through Topochemical K-deintercalation**

Electronic band structure calculations were performed for K$_x$Ni$_4$S$_2$ ($0 \leq x \leq 1$) with different K-deintercalation levels (**Fig.2**, see details in Supporting information). Two main features are noteworthy across the whole series. The first remarkable feature is the presence of Dirac cones, a signature of topological materials,[28-31] from **Γ-X** and **Z-A** points above the Fermi level. The Fermi level sits right below Dirac cones for x = 1 (**Fig.2a**, E$_{Dirac}$ = 25 meV for x = 1), and gradually shifts away with reduced degeneracy as the K-deintercalation proceeds (**Fig.2b-2c**, E$_{Dirac}$ = 259 meV for x = 0). The splitting of the Dirac cone along **Γ-X** for the x = 0 compound is likely a result of the unit cell change caused by the shifting of the Ni$_4$S$_2$ layers induced by the K-deintercalation;[53] however, it is interesting that the Dirac cone degeneracy remains the same along **Z-A**, suggesting a different local electronic environment between the two Ni-Ni square net layers.

Besides Dirac cones, another interesting characteristic of the electronic structure involves multiple flat bands below the Fermi level near the **M**-point along **X-M-Γ**. Opposite to the evolution of Dirac cones, the Fermi level resides distantly above flat bands for the x = 1 compound (**Fig.2a**, E$_{flat}$ = -239 meV for x = 1), and the K-deintercalation progressively shifts the Fermi level towards flat bands (**Fig.2b-2c**, E$_{flat}$ = **-**82 meV for x = 0). The shift of the Fermi level upon K-deintercalation is corroborated by the work function measurement along with the discovery of largely enhanced interlayer interactions between adjacent Ni$_2$S layers during the K-deintercalation (**Fig.S1-S2**).

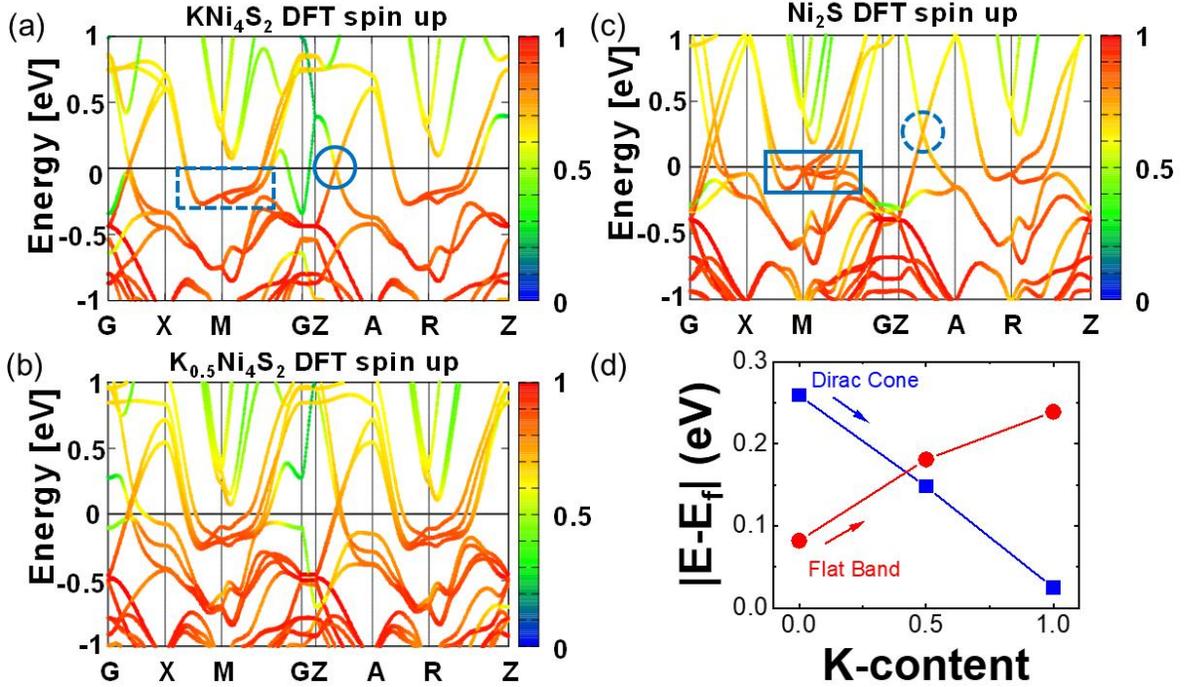

**Figure 2. Evolution of ground states revealed from electronic band structure calculations.** Density functional theory calculations of $K_xNi_4S_2$ with band structures showing projections of Ni $3d$ electrons for **(a)** $KNi_4S_2$ (x = 1). Note: Dirac cones are visible at the Fermi level between the G-X and Z-A points. **(b)** $K_{0.5}Ni_4S_2$ (x = 0.5) and **(c)** $Ni_2S$ (x = 0). The scale on the right side of each panel illustrates the fractional contributions of Ni $3d$ electrons. Flat bands are visible near the Fermi level at the M point; Note the flat bands and Dirac cones are highlighted with a square and circle, respectively, with solid lines indicating proximity to the Fermi surface and dashed lines representing away from the Fermi surface; **(d)** a summary of $|E-E_f|$ as a function of K-content for the Dirac cone and Flat band features.

Further topological invariant analysis, including the Wannier charge center calculations (**Fig.S3**), confirms the nontrivial $Z_2$ invariant of $KNi_4S_2$ (x = 1). The $Z_2$ number at all $k_i = 0.0$ ($i = 1, 2, 3$ for $x, y, z$) planes are 1, whereas those at $k_i = 0.5$ ($i = x, y, z$) planes are 0. These findings indicate that $KNi_4S_2$ (x = 1) is a strong topological material with a $Z_2$ index of 1;(000) according to the Fu-Kane classification,[62] similar to that of $Bi_2Se_3$.[63] Combined with the observed band inversion near the Fermi level, $KNi_4S_2$ (x = 1) is expected to be a topologically nontrivial Dirac metal. In contrast, a trivial electronic structure is observed for the fully deintercalated $Ni_2S$ (x = 0), both in nonmagnetic and antiferromagnetic spin configurations, with a trivial $Z_2$ index of 0;(000). The band inversion is also observed in the nonmagnetic configuration of $Ni_2S$ (x = 0), however, the inverted band is far away from the Fermi level, which results in trivial band topology (**Fig.S4**). Interestingly, such band inversion vanishes with the presence of an antiferromagnetic order, signifying that the topological feature in $K_xNi_4S_2$ may emerge as a synergistic consequence of magnetic order and charge carrier adjustments into the $Ni_4S_2$ layers from K-deintercalation.

In essence, our first-principles calculations (**Fig.2d**) predict a tunable ground state of $K_xNi_4S_2$ with the Dirac-cones-dominating state on the x = 1 ($KNi_4S_2$) side, extending towards the flat-bands-dominating state on the fully K-intercalated x = 0 ($Ni_2S$) side. In the following section, we provide our experimental evidence to showcase the rich and tunable ground states of $K_xNi_4S_2$ ($0 \leq x \leq 1$).

**Topological Dirac Metal**

A linearly-temperature-dependent resistivity is observed across all $K_xNi_4S_2$ compositions (**Fig.3a**), despite approaching the Ioffe-Regel limit upon K-deintercalation,[64] suggesting the strange metal behavior. The persistence of such non-Fermi liquid behavior highlights the underlying strong correlations across the $K_xNi_4S_2$ system[12,52,56,65,66] and the interplay between Dirac cones and flat bands.[60] In the following, we begin by examining the K-rich region, which projects the physical properties induced by the Dirac cones.

A signature of Dirac metals is the linearly dispersed band near the Fermi level, leading to the high mobility of charge carriers. To explore such characteristics, we performed magnetoresistance (defined as MR = [ρ(H)-ρ(0)]/ρ(0) × 100%) utilizing a horizontal rotator with the magnetic field rotating between the out-of-plane (θ = 90°) and in-plane direction (θ = 0°), relative to the $Ni_2S$ plane of the single crystal under investigation. A single crystal with a high K-content of x = 0.7 was first measured. The anisotropic magnetoresistance aligns with the low-dimensional nature of the system (**Fig.3b**), whereas the large magnetoresistance suggests the existence of high-mobility charge carriers near the Fermi surface. It is worth mentioning that non-saturating linear magnetoresistance (**Fig.3c, Fig.S7c**) bears similarities to many high-$T_c$ cuprates, and the coexistence with the strange metal behavior (**Fig.3a**) further highlights the presence of strong correlations.[56]

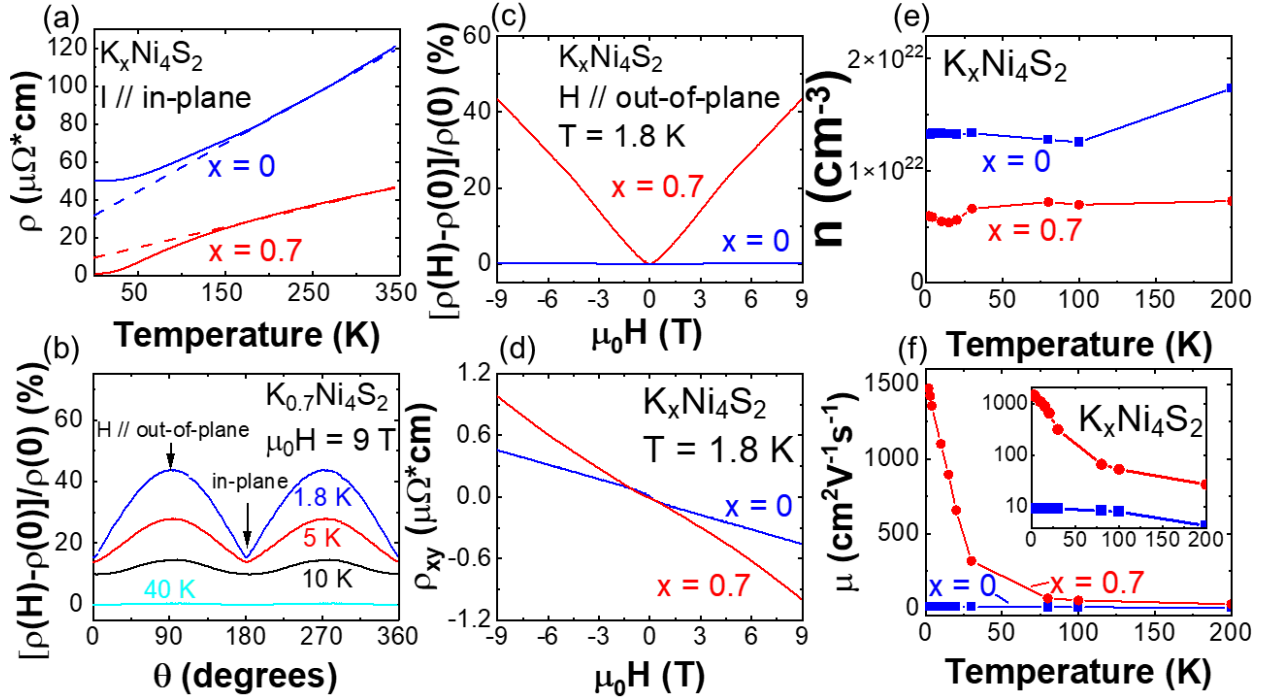

**Figure 3. Dirac-cones-dominating states** revealed by the electrical transport of $K_xNi_4S_2$, with **(a)** temperature-dependent in-plane electrical resistivity; **(b)** angular-dependent magnetoresistance of $K_{0.7}Ni_4S_2$ with 9 T of magnetic field; comparisons of **(c)** magnetoresistance and **(d)** Hall effect at 1.8 K between x = 0.7 and x = 0; the extracted **(e)** carrier density and **(f)** carrier mobility as a function of temperature. **(f) inset** shows the carrier mobility on a log scale.

To demonstrate the shift of the Fermi level away from Dirac cones induced by the K-deintercalation, we performed similar measurements on another single crystal with fully deintercalated K-content (x = 0). The diminishing magnetoresistance of the x = 0 compound confirms a significantly reduced carrier mobility due to the shifting away of Dirac cones (**Fig.3c** and **Fig.S8**). As suggested by the Hall effect data, electrons are the dominant carriers across the whole series of $K_xNi_4S_2$ (**Fig.3d** and **Fig.S9**), with the x = 0 compound showing a two-fold enhancement of carrier density compared to the x = 0.7 compound (**Fig.3e**), likely originating from flat bands near the Fermi surface. Extracted from the Hall effect data, a quantitative comparison shows an order-of-magnitude carrier mobility reduction from $\mu$ (x = 0.7) = 1471 $cm^2V^{-1}s^{-1}$ down to $\mu$ (x = 0) = 9.4 $cm^2V^{-1}s^{-1}$ at 1.8 K can be achieved through the K-deintercalation process (**Fig.3f** and **Fig.S10**), which is consistent with the DFT prediction discussed previously (**Fig.2**). The electrical transport results support the first-principles predictions that the Fermi level shifts away from Dirac cones as the K-deintercalation process proceeds.

We now focus on the emergence of the flat band near the Fermi level as the system approaches the more K-deintercalated region, starting with the heat capacity results, followed by flat band magnetism.

**Flat-Bands-Induced Magnetism**

The low-temperature heat capacity of $K_xNi_4S_2$ can be well described by $C(T) = \gamma T+\beta T^3$, where $\gamma$ and $\beta$ are the Sommerfeld coefficient and Debye constant, respectively (**Fig.4a-4b**). The Sommerfeld coefficient, $\gamma$, represents the electronic contribution. The flatness of the electronic bands near the Fermi surface can be evaluated through $\gamma$, as a large value of $\gamma$ typically suggests enhanced electronic correlations, for example, due to the flat band classically seen in the heavy fermion systems.[11] As expected, a more than two-fold enhancement of $\gamma$ is observed as the K-deintercalation process proceeds, increasing from $\gamma = 32.9$ mJ/mole/$K^2$ for x = 1 to $\gamma = 57.92$ mJ/mole/$K^2$ for x = 0.7 first, and eventually to $\gamma = 75.99$ mJ/mole/$K^2$ for x = 0 (**Fig.4c**). Although a large value of $\gamma$ is observed across the whole $K_xNi_4S_2$ system, the Kadowaki-Woods ratio analysis suggests that the $K_xNi_4S_2$ is a *d*-band metal instead of a heavy fermion material (**Supplemental Note S2** and **Fig.S6**). Nevertheless, the enhanced electronic correlations suggested by the $\gamma$ values corroborate the first-principles predictions that the Fermi level moves towards flat bands as the K-deintercalation process proceeds (**Fig.2d**).

The enhanced electronic correlations originating from flat bands can favor electronic instabilities, such as ferromagnetism, known as Stoner's criterion for ferromagnetism.[7] Accordingly, an emerging antiferromagnetic transition below 10 K gradually develops as the K-deintercalation proceeds in $K_xNi_4S_2$ (**Fig.4d-4g** and **Fig.S11**). However, both the anomalously large temperature-independent term $\chi_0$ (**Fig.S11a-b**) and the existence of ferromagnetic-like signals from isothermal magnetization, apparent even well above $T_N$ (**Fig.S11c**), suggest the presence of small ferromagnetic impurities with high $T_c$. Such impurity is attributed to metallic Ni that forms in the strong basic synthesis environment.[53] Among all the samples characterized, the x = 0.7 shows the least, if not absence of ferromagnetic impurity, thus, we use the x = 0.7 as the baseline to normalize the rest of the data by a constant $\chi_0$. A detailed discussion regarding the removal of the ferromagnetic impurity signal is included in the **Supplemental Note S4**.

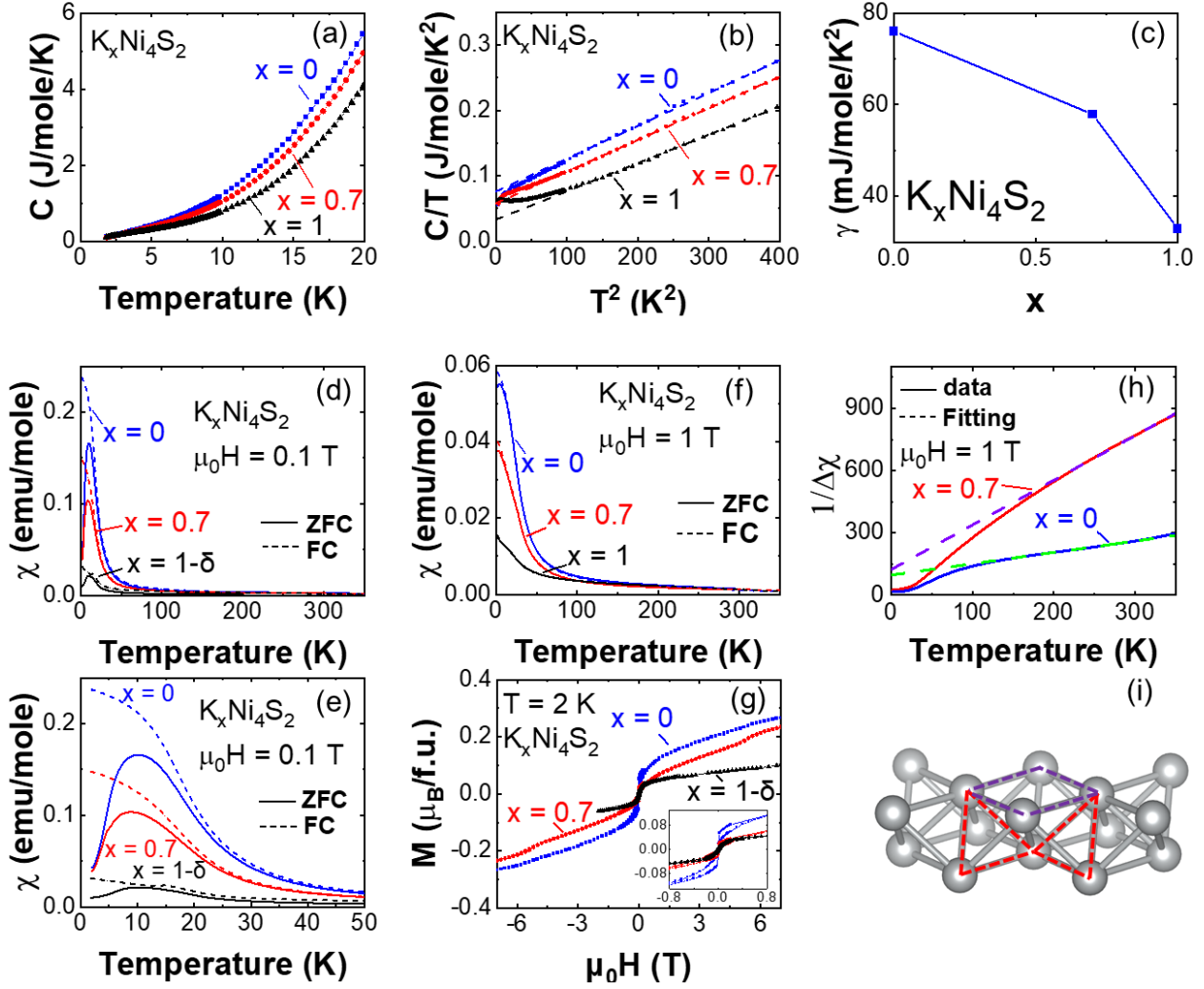

**Figure 4. Flat-bands-dominating properties** evidenced from heat capacity and magnetic susceptibility measurements of $K_xNi_4S_2$, with **(a)** C Vs. T and **(b)** C/T Vs. $T^2$ for x = 0, 0.7, and 1; **(c)** the extract Sommerfeld coefficient γ. **(d)** magnetic susceptibility with 0.1T magnetic field following Field-Cool (FC) and Zero-Field-Cool (ZFC) protocol; **(e)** a zoomed-in view at low-temperature region; **(f)** with 1 T magnetic field; **(g)** isothermal magnetization at T = 2 K with zoomed-in view as **(g) inset**; **(h)** the $1/\Delta\chi$ and the Curie-Weiss fitting; and **(i)** a schematic showing the local triangular (red dashed lines) and square geometry (purple dashed lines) of Ni that potentially leads to the observed spin frustrations.

After correcting for the impurity contribution, it is clear that an antiferromagnetic (AFM) order gradually develops upon K-deintercalation (**Fig.4d**), with a varying $T_N$ raised from 8.6 K for x = 0.7 to 10.1 K for x = 0 (**Fig.4e**). A transition is also observed in x = 1 with $T_N$ ~ 10 K. The substantially weaker signal observed in the x = 1 specimen originates from minor K-deintercalated phases formed during the sample washing process. Therefore, we label it as x = 1-δ. To test this interpretation, we conducted neutron powder diffraction (NPD) measurements on the x = 1 sample and observed no additional magnetic reflections between 260 K and 5 K, confirming the absence of long-range magnetic order. (**Fig.S12**). A weak magnetic hysteresis also gradually develops upon K-deintercalation (**Fig.4g inset**), suggesting a canted antiferromagnetic order. However, no spin saturation is observed up to 7 T (**Fig.4g**). The canted-AFM nature is further evidenced by the splitting of the magnetic susceptibility between the Field-cooling (FC) and Zero-field-cooling (ZFC) measurements, as well as the suppressed $T_N$ with increasing magnetic field (**Fig.4e-4f**, and **Fig.S13**). Alternatively, the bifurcation of FC and ZFC data may also be attributed to the spin glass state as the heat capacity reveals no clear 2$^{nd}$ order phase transition across the $T_N$ (**Fig.4a**). Hence, to clarify the magnetic ground state further, we carried out AC magnetic susceptibility at varying frequencies between 1 to 1000 Hz on the x = 0 compound. The lack of any frequency-dependent shift in either of the real ($\chi'$, **Fig.S14a**) and imaginary ($\chi''$, **Fig.S14b**) components across $T_N$ suggests the absence of spin-glass behavior. In addition, first-principles calculations also suggest competing ferromagnetic (FM) and AFM configurations are present on the x = 1 side, which could be another factor for the absence of magnetic order in $KNi_4S_2$. On the contrary, only one AFM configuration is preferred for the fully deintercalated x = 0 compound (**Fig.S15**).

For the Curie-Weiss analysis, we applied a strong magnetic field of 1 T to fully saturate the ferromagnetic impurity contribution, which was treated as a temperature-independent term, $\chi_0$. The inverse magnetic susceptibility above 200 K follows the Curie-Weiss law (**Fig.4h**), yielding a Curie–Weiss temperature, $\theta_{CW}$, and an effective moment, $\mu_{eff}$, of -57.4 K, 0.49 $\mu_B$/Ni and -177.7 K, 0.96 $\mu_B$/Ni for x = 0.7 and x = 0, respectively. The estimated effective moment is enhanced two-fold as the K-deintercalation proceeds. The effective moment is intermediate between $Ni^{1+}$ (1.73 $\mu_B$/Ni) and $Ni^0$, consistent with the electron-rich nature of $K_xNi_4S_2$, where the averaged valence state of Ni ranges from $Ni^{0.75+}$ (x = 1) to $Ni^{1+}$ (x = 0). The large negative $\theta_{CW}$ indicates strong antiferromagnetic interactions, consistent with the material's antiferromagnetic ground state. Furthermore, the resultant frustration parameter ($f = |\theta_{CW}|/T_N$) enhanced from 6.7 to 17.6 as the K-deintercalated from x = 0.7 to x = 0, indicating enhanced magnetic frustration[67] despite the precipitation of a magnetic order. The observed spin frustration may be attributed to the presence of local triangular geometry (red-dashed line in **Fig.4i**). In addition to the well-known geometric frustration associated with triangular geometry, frustration can also occur in square lattices (purple dashed line in **Fig. 4i**) when second-nearest-neighbor magnetic interactions dominate over those between nearest neighbors, a scenario observed in many Fe-based superconductors.[68-70] Interestingly, the coexistence of spin frustration, strange metal behavior, and flat-band-dominating Fermi surface closely parallels the characteristics of another kagome system, $Ni_3In$.[52]

In addition to flat-band-induced magnetism, the enhanced electron correlations may also lead to flat-band superconductivity, as discovered recently in twisted bilayer graphene.[1-4] Furthermore, the strange metal behavior and linear magnetoresistance of $K_xNi_4S_2$ (**Fig.3**) also resemble high-$T_c$ cuprates.[56] However, we observed no evidence of superconductivity down to 3.4 K under 10 GPa of hydrostatic pressure (**Fig.S16**). Instead, pressure-induced lattice distortions, as indicated by a reduced Residual-Resistance-Ratio ratio and orthorhombic distortions (**Figs.S16a-16b**), likely dominate the observed physical properties.

**Conclusion**

The $K_xNi_4S_2$ ($0 \leq x \leq 1$) simultaneously hosts both flat bands and Dirac cones at two different energies that can be accessed with intercalation/deintercalation processes and represents a unique case of a system lacking the Kagome or honeycomb lattice. The flexibility of hosting a wide range of K-content in its structure enables a change of the Fermi level over a wide range of energy. Notably, the ground state of $K_xNi_4S_2$ ($0 \leq x \leq 1$) can be fine-tuned from a non-magnetic topological Dirac metal ($KNi_4S_2$, x = 1) to a flat band containing antiferromagnetic metal ($Ni_2S$, x = 0) through a deintercalation process. In addition, our ex-situ topochemical K-deintercalation work demonstrates the feasibility for the potential in-situ control of quantum materials between Dirac-cones-dominating state, flat-band-dominating state, and intermediate states through, for instance, using electrochemical tuning of K-intercalation levels to move the Fermi energy.[71] This unique capability provides a platform for exploring fundamental physical phenomena and creates opportunities for applications, including reconfigurable electronics, multi-state memory devices, and adaptive sensors where dynamic control of electronic properties is necessary.

## Methods

**Synthesis.** For the synthesis of KNi$_4$S$_2$, Ni(OH)$_2$ and S were ground with a hydroxide flux consisting of a mixture of LiOH and KOH with the molar ratio of $n$(LiOH)/$n$(KOH) = 0.55/0.45 using a methodology described earlier.[72,73] The molar ratio of Ni(OH)$_2$:S was kept constant at about 1:2.5 to ensure an excess amount of S for the Ni source. Once all the starting materials were thoroughly mixed, they were loaded into a glassy carbon crucible and then placed in a quartz tube. The quartz tube was placed in a tube furnace horizontally with both ends connecting to O-ring-sealed metal flanges that allowed an inert gas, such as N$_2$, to pass through during the reaction. The furnace was heated to 450 °C at the rate of 5 °C/min, then kept at 450 °C for 20-30 hours, and then cooled to room temperature (RT) at a rate of 7-50 °C/hour (cooling rate appeared not to affect the product). After the furnace was cooled to room temperature (RT), the reaction vessel was immersed in a beaker filled with water and then ultrasonicated until all hydroxides were dissolved. The ternary K-Ni-S products were obtained by decantation of the solution and subsequent washing with methanol and drying in air. We found two methods for controllable K-deintercalation: **1)** using a flux composition that favors a Ni oxidation state higher than Ni$^{0.75+}$, and **2)** synthesizing the KNi$_4$S$_2$ first and soaking KNi$_4$S$_2$ crystals with hydroxide fluxes in methanol. We found that the K-deintercalation level is more homogeneous and controllable using the second method. The level of K-deintercalation was controlled by soaking time while washing the residual hydroxide flux with methanol. All ternary products appeared to be stable in the air for up to a few days. Crystals up to 1 mm in length could be obtained and used for subsequent structural determination and physical properties characterization.

**X-ray and neutron diffraction at ambient pressure.** Powder X-ray Diffraction (PXRD) data were collected using a Rigaku Miniflex X-ray diffractometer with Cu K$_\alpha$ radiation, $\lambda$ = 1.5418 Å. Temperature-dependent single crystal X-ray diffraction (SXRD) between 100 - 300 K was collected using synchrotron x-ray with $\lambda$ = 0.41328 Å at 15-IDD at the Advanced Photon Source (APS). Neutron powder diffraction (NPD) measurements were performed using the BT-1 high-resolution neutron powder diffractometer at the NIST Center for Neutron Research with the Ge (311) monochromator ($\lambda$ = 2.0772 Å) and the 60' in-pile collimator. In addition to base temperature measurement (~ 4 K), data at 25 K and 260 K were also collected to investigate possible structural or magnetic transitions.

**DC magnetic susceptibility ($\chi_{dc}$).** X$_{dc}$ data for coaligned single crystals were collected with a Magnetic Property Measurement System (MPMS3, Quantum Design). Temperature-dependent magnetic measurements were performed from 2 to 300 K in both field-cooled (FC) and zero-field-cooled (ZFC) conditions, while field-dependent data were collected under applied fields of up to 7 T.

**AC magnetic susceptibility ($\chi_{ac}$).** $\chi_{ac}$ was measured with a 14 T Quantum Design Physical Property Measurement System (PPMS-14) on powder samples. Zero field-cooled measurements were taken from 40 K to 1.8 K with varying AC-frequencies from 1 to 1000 Hz.

**Electrical Resistivity, Hall Effect and heat capacuty measurements.** Electrical resistivity with the standard four-probe method and Hall effect with standard four-probe Hall configuration were performed with a single crystal of K$_x$Ni$_4$S$_2$ using a 9 T Dynacool Quantum Design Physical Property Measurement System (PPMS-Dynacool) equipment with a horizontal rotator. Heat capacity measurements of coaligned single crystals of K$_x$Ni$_4$S$_2$ were carried out on a Physical Property Measurement System (PPMS-Dynacool).

**Theoretical methods.** We adopted the density functional theory (DFT) calculation to compute the band structure and the Fermi surface of KNi$_4$S$_2$, K$_{0.5}$Ni$_4$S$_2$, and Ni$_2$S using Vienna Ab-initio

Simulation Package (VASP).[74,75] The Perdew-Burke-Ernzerhof for solids (PBE-sol)[76] implementation was used as the exchange-correlation functional within DFT. The gamma-centered Monkhorst-Pack $k$-point mesh of 8×8×2 and the plane-wave energy cutoff of 400eV were used for both materials. Molecular orbital calculations were performed using Spartan'24 with a range-separated hybrid generalized gradient approximation (RSH-GGA) ωB97X-D as the functionals.[77] The electronic structure topological analysis was performed using WannierTools.[78] The Wannier basis and Hamiltonian were constructed using Wannier90[79] based on density functional theory calculations using Vienna ab-initio simulation package (VASP).[74,75] The band structure and density of states were also computed using VASP. Projector augmented-wave method and plane-wave basis were employed for electron orbitals, charge density, and electron-nuclei interaction with a cutoff energy of 400 eV.[80] Hubbard $U$ corrections for Ni $d$ orbitals were added in the electronic structure topological analysis to describe the orbital localization with U = 5 eV properly. Noncollinear magnetization with spin-orbit couplings was considered within the generalized gradient approximation and Perdew−Burke−Ernzerhof functional.[81]

**Electrical Transport Properties under High Pressure.** Electrical transport measurements of $KNi_4S_2$ bulk single crystals were conducted using the quasi-four-terminal method with NaCl as the pressure-transmitting medium. All the measurements were carried out in a diamond anvil cell (DAC) with a 500 μm culet size. The gasket was prepared from a 250 μm-thick stainless steel sheet, which was pre-indented to a thickness of approximately 60 μm. For all measurements, an initial hole of ~150 μm in diameter was created. The indented gasket was then insulated using a fine alumina and epoxy mixture. After insulation, a ~120 μm hole was laser-drilled through the compressed gasket. Fine NaCl powder was loaded into the hole, and pressure was gradually increased to ~2.0 GPa until the NaCl turned transparent. Then, a rectangular plate-like $KNi_4S_2$ single crystal was carefully loaded on top of the transparent NaCl, and four platinum electrodes were arranged to contact the sample. The temperature-dependent resistance of $KNi_4S_2$ was measured in a closed-cycle cryocooler system (2.4–300 K) using a Stanford Research SR830 digital lock-in amplifier with a current of 100 - 1000 μA. Pressure was measured using the ruby fluorescence method,[82] both at room temperature and low temperature of ~4.2 K, with the aid of an online ruby system.

**Powder X-ray diffraction under high pressure.**
High-pressure X-ray diffraction (XRD) measurements were conducted using a symmetric diamond anvil cell (DAC) with 500 μm culet Boehler-type diamonds and a stainless steel gasket. Pressure was determined using both ruby fluorescence[83] and the equation of state of gold (Au), with measurements taken before and after each diffraction scan to ensure accuracy. A $KNi_4S_2$ powder sample, approximately 25 × 25 × 10 μm³ in size, was loaded into the DAC, with neon used as the pressure-transmitting medium (PTM). X-ray diffraction data were collected up to 20.2 GPa. The X-ray experiments were performed at beamline 16-BM-D of the High Pressure Collaborative Access Team (HPCAT) at the Advanced Photon Source (APS), Argonne National Laboratory (ANL), using X-ray with a wavelength of λ = 0.4066 Å. Diffraction images were collected in angle-dispersive geometry while rotating the DAC by approximately 50°, enabling comprehensive capture of the Bragg peaks. The 2D images were integrated using the Dioptas program,[84] and the resulting powder patterns were analyzed by Le Bail fitting using the GSAS-EXPGUI package.[85] Single-crystal X-ray structures of $KNi_4S_2$ were used as starting models for the refinement of all the powder data. Although data quality declined above 12.2 GPa, the position of the diffraction peaks indicated no structural phase transition up to the maximum pressure measured (20.2 GPa).

Moreover, the least-squares fitting of *P-V* data from the third-order Birch-Murnaghan equation of state (EoS) was performed with the aid of EoSFit7c software.[86]

**Scanning Transmission Electron Microscopy (SEM/TEM) characterization.** Microscopic images were examined on a Hitachi SU-70 SEM field emission scanning electron microscope (SEM), and their elemental compositions were determined by energy dispersive X-ray spectroscopy (EDS) using a BRUKER EDS detector.

**Photoemission Yield Spectroscopy in Air (PYSA):** PYSA was measured using a Riken-Keiki AC-2 instrument equipped with a tunable monochromatic ultraviolet (UV) light source that produces light in the 4.2 – 7 eV range. The $K_xNi_4S_2$ ($x$ = 0, 0.7, 1) materials were scanned in the 4.2 – 6.2 eV range, with the generated photoelectrons being measured at each energy step. $K_0Ni_4S_2$ (x = 0) was measured using a 0.05 eV step size and 20-second counting time with a UV intensity of 100 nW. The data for $K_{0.7}Ni_4S_2$ (x = 0.7) and $KNi_4S_2$ (x = 1) were collected using a 0.1 eV step size and 10-second counting time with a UV intensity of 800 nW. For all samples, a quantity of light correction was utilized over the measured energy range to account for fluctuations in the UV lamp's intensity. The square root of the photoelectron counts per second was plotted versus the energy of the incident light, and the work function of each sample was determined through the intersection of the regression lines from the baseline and linear onset of the obtained PYSA data. The negative of the work function value provides information on the position of the Fermi level of the material relative to the free electron.

**Resource availability**

**Lead Contact**

Requests for further information and resources should be directed to and will be fulfilled by the lead contact, Mercouri G. Kanatzidis (m-kanatzidis@northwestern.edu).

**Materials availability**

The materials generated in this study may be made available upon request.

**Data and code availability**

• All data reported in this paper, such as the experimental section, additional physical property characterizations such as magnetotransport, heat capacity, magnetic susceptibility measurements, and photoemission yield spectroscopy in air, will be shared by the lead contact upon request.

• This paper does not report original code.

• Crystallography data have been deposited at the Cambridge Crystallographic Data Centre under the database identifier CCDC 2088104−2088106 and are publicly available as of the date of publication.

**SUPPLEMENTAL INFORMATION**

Supporting information can be found online at . https://doi.org/10.1016/j.matt.2025.102418.

**Acknowledgments**

This work is primarily supported by the U.S. Department of Energy, Office of Science, Basic Energy Sciences, Materials Sciences, and the Engineering Division. This material is based upon work supported by Laboratory Directed Research and Development (LDRD) funding from Argonne National Laboratory, provided by the Director, Office of Science, of the U.S. Department of Energy under Contract No. DE-AC02-06CH11357. The work (SEM/EDX) performed at the Center for Nanoscale Materials, a U.S. Department of Energy Office of Science User Facility, was supported by the U.S. DOE, Office of Basic Energy Sciences, under Contract No. DE-AC02-06CH11357. Work at the beamline 15-IDD at the Advanced Photon Source (APS) at Argonne National Laboratory was supported by the U.S. Department of Energy, Office of Science, Office of Basic Energy Sciences under Contract No. DE-AC02-06CH11357. NSF's ChemMatCARS Sector 15 is supported by the Divisions of Chemistry (CHE) and Materials Research (DMR), National Science Foundation, under grant no. NSF/CHE-1834750 and NSF/CHE-2335833. X.Z. acknowledges support from the Georgetown University startup fund. T.D. acknowledges the computational resources provided by the National Supercomputer Center in Tianjin. H. Park acknowledges the computing resources provided by Bebop, a high-performance computing cluster operated by the Laboratory Computing Resource Center at Argonne National Laboratory. The research at the University of Utah is supported by the National Science Foundation Division of Materials Research Award No. 2132692 as well as the U.S. Department of Energy (DOE) Office of Science, Fusion Energy Sciences funding award entitled High Energy Density Quantum Matter, Award No DE-SC0020340. R. Kumar and R. J. Hemley acknowledge support from DOE-NNSA (DE-NA0003975) through the Chicago/DOE Alliance Center and DOE-SC (DE-SC0020340). HPCAT operations are supported by the DOE-NNSA Office of Experimental Sciences. The Advanced Photon Source is a DOE Office of Science User Facility operated for the DOE Office of Science by Argonne National Laboratory under Contract No. DE-AC02-06CH11357.

**Author Contributions**

The work was conceived by H.Z., X.Z., D.Y.C., and M.G.K., with input from all authors. H.Z. carried out the synthesis, X-ray diffraction, electrical transport, heat capacity, DC magnetic susceptibility measurements, data analysis, and writing. X.Z. carried out the powder neutron diffraction, synchrotron X-ray diffraction, molecular orbital bonding analysis, and writing. H.P.

performed first-principles electronic band structure calculations. T.D. analyzed topological properties and contributed to writing. B.W. performed AC magnetic susceptibility measurements. C.M.B. and H.W. conducted powder neutron diffraction measurements. S.E.P. and Z.-L.X. contributed magnetotransport data analysis. T.B. performed high-pressure transport measurements and high-pressure XRD analysis. T.M. performed high-pressure transport measurements. W. C. analyzed high-pressure XRD results. R. K. collected and analyzed high-pressure XRD results. S.D. performed high-pressure electrical resistivity measurements. R. H. contributed to high-pressure powder diffraction measurements. A.P.A, II conducted and analyzed photoemission yield spectroscopy in air (PYSA) measurements. Y.-S. C collected synchrotron X-ray diffraction. S.R., D.Y.C., and M.G.K. supervised the project.

**Conflicts of interest**

The authors declare no competing financial interests.

**Note**

Certain trade names and company products are identified in order to specify adequately the experimental procedure. In no case does such identification imply recommendation or endorsement by the National Institute of Standards and Technology, nor does it imply that the products are necessarily the best for the purpose.